\pdfoutput=1

\documentclass{article}
\usepackage{ulem}  
\usepackage{graphicx}
\begin{document}
\input epsf.tex    
\input epsf.def   

\input psfig.sty


\title{Statistical methods for cosmological parameter selection and estimation}

\markboth{Andrew R.\ Liddle}{Cosmological parameters}

\author{Andrew R.\ Liddle\\
Astronomy Centre, University of Sussex, Brighton BN1 9QH, UK}


\maketitle

\begin{abstract}
The estimation of cosmological parameters from precision observables is an
important industry with crucial ramifications for particle physics. This article discusses the statistical methods presently used in cosmological data analysis, highlighting the main assumptions and uncertainties. The topics covered are parameter estimation, model selection, multi-model inference, and experimental design, all primarily from a Bayesian perspective.
\end{abstract}

\section{INTRODUCTION}

During the last decade, cosmology has advanced from an era of largely qualitative questions --- is the Universe flat, open or closed?, does dark energy exist in the Universe?, etc. --- to one where precision determinations of many of the Universe's properties are possible. We have cosmological models capable of explaining the detailed observations available, and whose parameters are beginning to be pinned down at the ten percent, and in some cases one percent, level \cite{WMAP5}. Nevertheless, quality cosmological data are an expensive resource and it is imperative to make the best possible use of them. This implies use of the best available statistical tools in order to obtain accurate and robust conclusions.

For around a decade now, the established leading cosmological model considers five material constituents: baryons (taken, imprecisely, to include electrons), photons, neutrinos, cold dark matter (CDM), and dark energy. The simplest model for dark energy, a cosmological constant $\Lambda$, is in excellent agreement with observations, and the model is then known as a $\Lambda$CDM model. The most important constraints come from the evolution of cosmic structures. These are seeded by small initial density perturbations, which in the standard cosmological model are taken as adiabatic, gaussian, and nearly scale-invariant, as predicted by the simplest models of cosmological inflation \cite{LidLyth}. 

\begin{figure}[t]
\centering
\includegraphics[width=0.8 \linewidth]{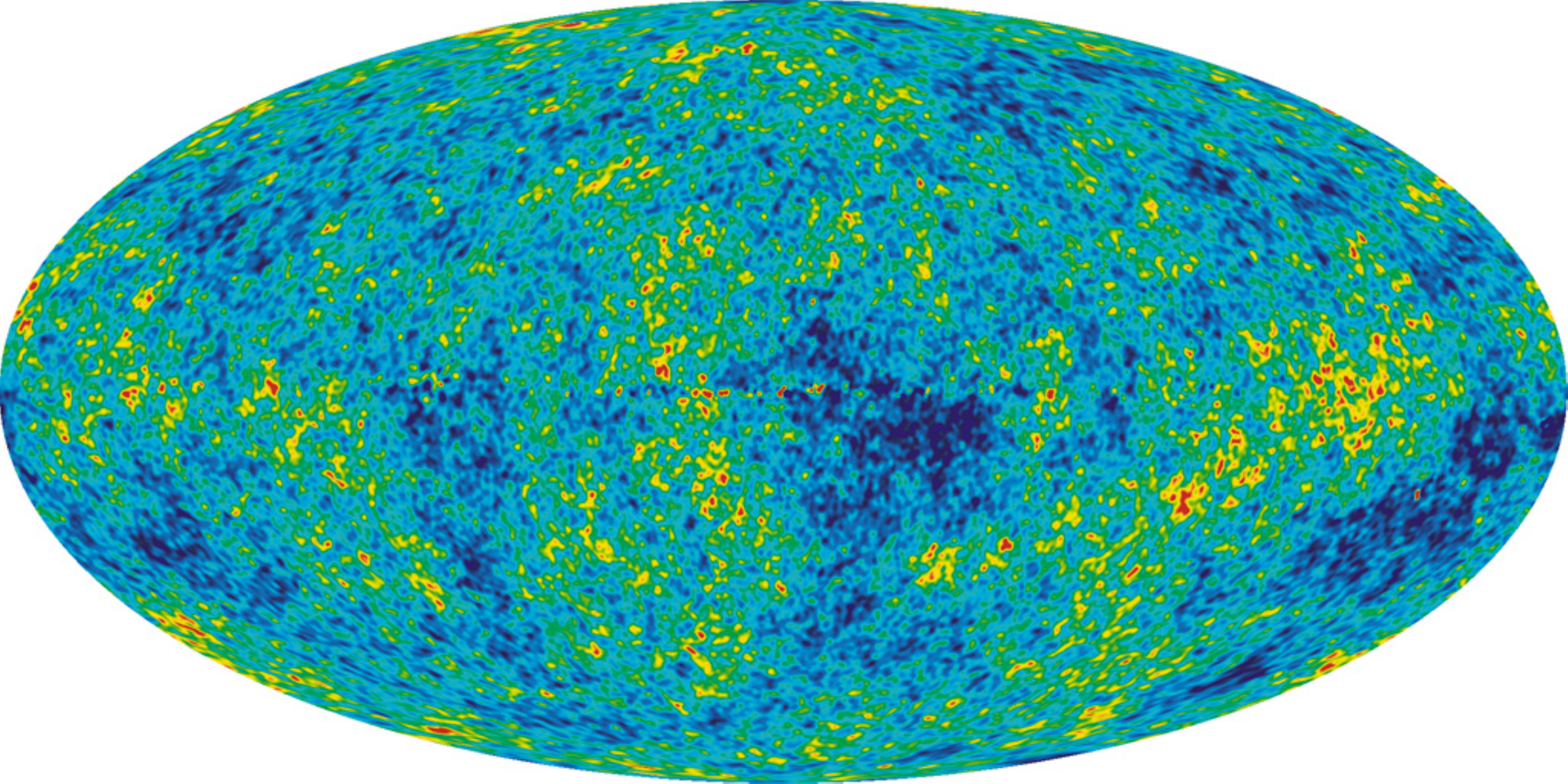}
\caption{Cosmic microwave background anisotropies as imaged by WMAP from five years of accumulated data. [Figure courtesy NASA/WMAP Science Team.]}
\label{f:liddle1}
\end{figure}

This model is supported and constrained by a series of cosmological observations. Most important are measurements of cosmic microwave background (CMB) anisotropies, particularly by the Wilkinson Microwave Anisotropy Probe (WMAP) as shown in Figure~\ref{f:liddle1}. Typical analyses also incorporate other data, such as galaxy clustering data, the luminosity distance--redshift relation of Type Ia supernovae, and direct measures of the Hubble constant. The region of parameter space where the $\Lambda$CDM model matches all those data is often referred to as the concordance model.

In its very simplest incarnation, the photon density is taken to be well measured by the CMB temperature, neutrinos to be nearly massless, and the Universe spatially flat. The model then features only four fundamental parameters: the Hubble parameter $h$, the densities of baryons $\Omega_{\rm b}$ and CDM $\Omega_{\rm CDM}$, and the amplitude of primordial density perturbations $A_{\rm S}$. In addition, a comparison with data will usually include extra phenomenological parameters, which in principle can be computed from the above but which in practice cannot be reliably. For cosmic microwave background studies, the optical depth $\tau$, measuring the fraction of CMB photons scattering from ionized gas at low redshift, is needed, while use of galaxy clustering data may require inclusion of the galaxy bias parameter $b$ which relates galaxy clustering to dark matter clustering. 

Beyond those basic parameters, cosmologists hope that future investigations will uncover new physical processes, permitting extra parameters to be incorporated and measured. In some cases, it is more or less certain that the parameter is relevant and only a matter of time before observational sensitivity becomes sufficient. Examples here are neutrino masses and the cosmic Helium fraction (though the latter is again in principle computable from other parameters, independent verification of its value would be an important consistency check). 

Much more numerous, though, are parameters describing effects which may or may not be relevant to our Universe. An extensive list is given, for instance, in Ref.~\cite{lid04}. Are the primordial perturbations precisely scale invariant, or do they have a scale dependence quantified by the spectral index $n$? Do primordial gravitational waves exist, as predicted by inflation? Does the dark energy density evolve with time? Are there cosmic strings in the Universe? Are the initial perturbations really adiabatic and gaussian? A fuller account of these parameters can be found for instance in Ref.~\cite{LahavLid}.

In summary, creation of precision cosmological models is an ongoing process with two distinct goals. One is to determine the set of parameters, i.e.\ physical processes, necessary to describe the available observations. The second is to determine the preferred values of these parameters.  We can then pursue the ultimate aim of relating cosmological observations to underlying fundamental physics.

\section{INFERENCE}

\subsection{Orientation}

Inference is the method by which we translate experimental/observational information into constraints on our mathematical models. The model is a representation of the physical processes that we believe are relevant to the quantities we plan to observe. To be useful, the model must be sufficiently sophisticated as to be able to explain the data, and simple enough that we can obtain predictions for observational data from it in a feasible time.  At present these conditions are satisfied in cosmology, with the best models giving an excellent representation of that data, though the computation of theoretical predictions for a representative set of models is a supercomputer class problem. Particularly valued are models which are able to make distinctive predictions for observations yet to be made, though Nature is under no obligation to behave distinctively.

The data which we obtain may be subject only to experimental uncertainty, or they may also have a fundamental statistical uncertainty due to the random nature of underlying physical processes. Both types of data arise in cosmology. For instance, the present expansion rate of the Universe (the Hubble constant), could in principle be measured to near-arbitrary accuracy with sufficiently advanced instrumentation. By contrast, the detailed pattern of cosmic microwave anisotropies, as measured by WMAP, is not believed to be predictable even in principle, being attributed to a particular random realization of quantum processes occurring during inflation \cite{LidLyth}. Observers at different locations in the Universe see different patterns in the CMB sky, the cosmological information being contained in statistical measures of the anisotropies such as the power spectrum. Observers at any particular location, such as ourselves, can witness only our own realization and there is an inherent statistical uncertainty, {\em cosmic variance}, that we cannot overcome, but which fortunately can be modelled and incorporated in addition to measurement uncertainty.

A model will typically not make unique predictions for observed quantities; those predictions will instead depend on some number of {\em parameters} of the model. Examples of cosmological parameters are the present expansion rate of the Universe, and the densities of the various constituents such as baryons, dark matter, etc. Such parameters are not (according to present understanding, anyway) predictable from some fundamental principle; rather, they are to be determined by requiring that the model does fit the data to hand. Indeed, determining the values of such parameters is often seen as the primary goal of cosmological observations, and Chapter~\ref{c:cospar} is devoted to this topic.

At a more fundamental level, several different models might be proposed as explanations of the observational data. These models would represent alternative physical processes, and as such would correspond to different {\em sets} of parameters that are to be varied in fitting to the data. It may be that the models are nested within one another, with the more complex models positing the need to include extra physical processes in order to explain the data, or the models may be completely distinct from one another. An example of nested models in cosmology is the possible inclusion of a gravitational wave contribution to the observed CMB anisotropies. An example of disjoint models would be the rival explanations of dark energy as caused by scalar field dynamics or by a modification to the gravitational field equations. Traditionally, the choice of model to fit to the data has been regarded as researcher-driven, hopefully coming from some theoretical insight, with the model to be validated by some kind of goodness-of-fit test. More recently, however, there has been growing interest in allowing the data to distinguish between competing models. This topic, {\em model selection} or model comparison, is examined in Chapter~\ref{c:modsel}.

The comparison of model prediction to data is, then, a statistical inference problem where uncertainty necessarily plays a role. While a variety of techniques exist to tackle such problems, within cosmology one paradigm dominates --- {\em Bayesian inference}. This article will therefore focus almost exclusively on Bayesian methods, with only a brief account of alternatives at the end of this section. The dominance of the Bayesian methodology in cosmology sets it apart from the traditional practice of particle physicists, though there is now increasing interest in applying Bayesian methods in that context (e.g.\ Ref.~\cite{Cowan}).

\subsection{Bayesian inference}

The Bayesian methodology goes all the way back to Thomas Bayes and his theorem, posthumously published in 1763 \cite{Bayes63}, followed soon after by pioneering work on probability by Laplace.  The technical development of the inference system was largely carried out in the first half of the 20th century, with Jeffreys' textbook \cite{Jeff} the classic source. For several decades afterwards progress was held up due to an inability to carry out the necessary calculations, and only in the 1990s did use of the methodology become widespread with the advent of powerful multiprocessor computers and advanced calculational algorithms. Initial applications were largely in the fields of social science and analysis of medical data, the volume edited by Gilks et al.~\cite{Gilks} being a particularly important source. 
The publication of several key textbooks in the early 21st century, by Jaynes \cite{Jaynes}, MacKay \cite{MacKay} and Gregory \cite{Gregory}, the last of these being particularly useful for physical scientists seeking to apply the methods, cemented the move of such techniques to the mainstream. An interesting history of the development of Bayesianism is given in Ref.~\cite{bayesianism}.

The essence of the Bayesian methodology is to assign probabilities to all quantities of interest, and to then manipulate those probabilities according to a series of rules, amongst which Bayes theorem is the most important. The aim is to update our knowledge in response to emerging data. An important implication of this set-up is that it requires us to specify what we thought we knew {\it before} the data was obtained, known as the \emph{prior probability}. While all subsequent steps are algorithmic, the specification of the prior probability is not, and different researchers may well have different views on what is appropriate. This is often portrayed as a major drawback to the Bayesian approach. I prefer, however, to argue the opposite --- that the freedom to choose priors is the opportunity to express physical insight. In any event, one needs to check that one's result is robust under reasonable changes to prior assumptions.

An important result worth bearing in mind is a theorem of Cox \cite{Cox46}, showing that Bayesian inference is the unique consistent generalization of Boolean logic in the presence of uncertainty. Jaynes in particular sees this as central to the motivation for the Bayesian approach \cite{Jaynes}.

In abstract form, Bayes theorem can be written as
\begin{equation}
P(B|A) = \frac{P(A|B)P(B)}{P(A)} \,,
\end{equation}
where a vertical line indicates the conditional probability, usually read as `the probability of B given A'.
Here $A$ and $B$ could be anything at all, but let's take $A$ to be
the set of data $D$ and $B$ to be the parameter values $\theta$
(where $\theta$ is the $N$-dimensional vector of parameters being
varied in the model under consideration), hence writing
\begin{equation}
\label{e:bayes2}
P(\theta|D) = \frac{P(D|\theta)P(\theta)}{P(D)} \,.
\end{equation}

In this expression, $P(\theta)$ is the prior probability, indicating what we thought the probability of different values of $\theta$ was before we employed the data D. One of our objectives
is to use this equation to obtain the \emph{posterior probability} of the
parameters given the data, $P(\theta|D)$.  This is achieved by
computing the \emph{likelihood} $P(D|\theta)$, often denoted ${\cal L}(\theta)$ with the dependence on the dataset left implicit. 

\subsection{Alternatives to Bayesian inference}

The principal alternative to the Bayesian method is usually called the {\em frequentist approach}, indeed commonly a dichotomy is set up under which any non-Bayesian method is regarded as frequentist. The underpinning concept is that of sampling theory, which refers to the frequencies of outcomes in random repeatable experiments (often caricatured as picking coloured balls from urns). According to MacKay \cite{MacKay}, the principal difference between the systems is that frequentists apply probabilities only to random variables, whereas Bayesians additionally use probabilities to describe inference. Frequentist analyses commonly feature the concepts of {\em estimators} of statistical quantities, designed to have particular sampling properties, and {\em null hypotheses} which are set up in the hope that data may exclude them (though without necessarily considering what the preferred alternative might be).

An advantage of frequentist methods is that they avoid the need to specify the prior probability distribution, upon which different researchers might disagree. Notwithstanding the Bayesian point-of-view that one should allow different researchers to disagree on the basis of prior belief, this means that the frequentist terminology can be very useful for expressing results in a prior-independent way, and this is the normal practice in particle physics. 

A drawback of frequentist methods is that they do not normally distinguish the concept of a model with a fixed value of a parameter, versus a more general model where the parameter happens to take on that value (this is discussed in greater detail below in Section~\ref{c:modsel}), and they find particular difficulties in comparing models which are not nested.

\section{COSMOLOGICAL PARAMETER ESTIMATION}

\label{c:cospar}

\subsection{Goals and methodology}

In cosmological parameter estimation, we take for granted that we have a dataset D, plus a model with parameter vector $\theta$ from which we can extract predictions for those data, in the form of the likelihood ${\cal L}(\theta) \equiv P(D|\theta)$. Additionally, we will have a prior distribution for those parameters, representing our knowledge before the data was acquired. While this could be a distribution acquired from analyzing some previous data, more commonly cosmologists take the prior distribution to be flat, with a certain range for each parameter, and reanalyze from scratch using a compilation of all data deemed to be useful.

Our aim is then to figure out the parameter values which give the best fit to the data, or, more usefully, the region in parameter space within which the fit is acceptable, i.e.\ to find the maximum likelihood value and explore the shape of the likelihood in the region around it. In many cases one can hope that the likelihood takes the form of a multi-variate gaussian, at least as long as one doesn't stray too far from the maximum.

The task then is to find the high-likelihood regions of the function ${\cal L}(\theta)$, which sounds straightforward. However, there are various obstacles
\begin{itemize}
\item The likelihood function may be extremely sharply peaked, and it may have several maxima masquerading as the true maximum.
\item The parameter space may have a high dimensionality, cosmological examples often having 6 to 10 parameters independently varying.
\item There may be parameter degeneracies, where likelihood varies only weakly, or not at all, along some direction in parameter space.
\item The evaluations of the likelihood may be computationally demanding, either in generating the theoretical predictions from the model, or in computing the likelihood of those predictions. A typical likelihood evaluation in a cosmological calculation involving CMB anisotropies is a few seconds of CPU time.
\end{itemize}

In combination, these seriously obstructed early data analysis efforts, even when the dataset was fairly limited, because available computer power restricted researchers to perhaps $10^5$ to $10^6$ likelihood evaluations. Once beyond five or six parameters, which is really the minimum for an interesting comparison, brute-force mapping of the likelihood on a grid of parameters becomes inefficient, as the resolution in each parameter direction becomes too coarse, and anyway too high a fraction of computer time ends up being used in regions where the likelihood turns out to be too low to be of interest. 

This changed with a paper by Christensen and Meyer \cite{ChrisMey}, who pointed out that problems of this kind are best tackled by Monte Carlo methods, already extensively developed in the statistics literature, e.g.\ Ref.~\cite{Gilks}. Subsequently, Lewis and Bridle wrote the CosmoMC package \cite{LewBrid}, implementing a class of Monte Carlo methods for cosmological parameter estimation. The code has been very widely adopted by researchers, and essentially all cosmological parameter estimation these days is done using one of a variety of Monte Carlo methods.

\subsection{Monte Carlo methods}

\label{ss:mcmeth}

Monte Carlo methods are computational algorithms which rely on random sampling, with the algorithm being guided by some rules designed to give the desired outcome. An important subclass of Monte Carlo methods are {\em Markov Chain Monte Carlo (MCMC) methods}, defined as those in which the next `step' in the sequence depends only upon the previous one. The sequence of steps is then known as a Markov chain. Each step corresponds to some particular value of the parameters, for which the likelihood is evaluated. The Markov chain can therefore be viewed as a series of steps (or jumps) around the parameter space, investigating the likelihood function shape as it goes. 

You may find it convenient to visualize the likelihood surface as a mountainous landscape with one dominant peak. The simplest task that such a process could carry out would be to find the maximum of the likelihood: choose a random starting point, propose a random jump to a new point, accept the jump only if the new point has a higher likelihood, return to the proposal step and repeat until satisfied that the highest point has been found. Even this simple algorithm obviously needs some tuning: if the steps are too large, the algorithm may soon find it difficult to successfully find a higher likelihood point to jump to, whereas if they are small the chain may get stuck in a local maximum which is not the global maximum. That latter problem may perhaps be overcome by running a series of chains from different starting points.

Anyway, the maximum itself is not of great interest; what we want to know is the region around the maximum which is compatible with the data. To do this, we desire an algorithm in which the Markov chain elements correspond to random samples from the posterior parameter distribution of the parameters, i.e.\ that each chain element represents the probability that those particular parameter values are the true ones. The simplest algorithm which achieves this is the Metropolis--Hastings algorithm, which is a remarkably straightforward modification of the algorithm described in the previous paragraph.

\subsubsection{Metropolis--Hastings algorithm}

The Metropolis--Hastings algorithm is as follows:
\begin{enumerate}
\item Choose a starting point within the parameter space.
\item Propose a random jump. Any function can be used to determine the probability distribution for the length and direction of the jump, as long as it satisfies the `detailed balance' condition that a jump back to the starting point is as probable as the jump away from it. This is most easily done using a symmetric proposal function, e.g.\ a multivariate gaussian about the current point. Evaluate the likelihood at the new point, and hence the probability by multiplying by the prior at that point. [If the prior is flat, the probability and likelihood become equivalent.]
\item If the probability at the new point is higher, accept the jump. If it is lower, we accept the jump with a probability given by the ratio of the probabilities at the new and old point. If the jump is not accepted, we stay at the same point, creating a duplicate in the chain.
\item Repeat from Step 2, until satisfied that the probability distribution is well mapped out. This may be done for instance by comparing several chains run from different starting points, and/or by using convergence statistics amongst which the Gelman--Rubin test \cite{gelman,Gilks} is the most commonly used.
\end{enumerate}

By introducing a chance of moving to a lower probability point, the algorithm can now explore the shape of the posterior in the vicinity of the maximum. The generic behaviour of the algorithm is to start in a low likelihood region, and migrate towards the high likelihood `mountains'. Once near the top, most possible jumps are in the downwards direction, and the chain meanders around the maximum mapping out its shape. Accordingly, all the likelihood evaluations, which is where the CPU time is spent, are being carried out in the region where the likelihood is large enough to be interesting. The exception is the early stage, which is not representative of the posterior distribution as it maintains a memory of the starting position. This `burn-in' phase is then deleted from the list of chain points. 

Although any choice of proposal function satisfying detailed balance will ultimately yield a chain sampling from the posterior probability distribution, in practice, 
as with the simple example above, the algorithm needs to be tuned to work efficiently. This is referred to as the convergence of the chain (to the posterior probability).
The proposal function should be tuned to the scale of variation of the likelihood near its maximum, and if the usual choice of a gaussian is made its axes should ideally be aligned to the principal directions of the posterior (so as to be able to navigate quickly along parameter degeneracies). Usually, a short initial run is carried out to roughly map out the posterior distribution which is then used to optimize the proposal function for the actual computation.\footnote{One can also help things along by using variables which respect known parameter degeneracies in the data, e.g.\ that the CMB data is particularly good at constraining the angular-diameter distance to the last-scattering surface through the position of the peaks.} The resulting acceptance rate of new points tends to be around 25\%.

The upshot of this procedure is a Markov chain, being a list of points in parameter space plus the likelihood/posterior probability at each point. A typical cosmological example may contain $10^4$ to $10^5$ chain elements, and some collaborations including WMAP make their chains public. There is usually some short-scale correlation of points along the chain due to the proposal function; some researchers `thin' the chain by deleting elements to remove this correlation though this procedure appears unnecessary. By construction, the elements correspond to random samples from the posterior, and hence a plot of the point density maps it out. 

The joy of having a chain is that marginalization (i.e.\ figuring out the allowed ranges of a subset of the full parameter set, perhaps just one parameter) becomes trivial; you just ignore the other parameters and plot the point density of the one you are interested in. By contrast, a grid evaluation of the marginalized posterior requires an integration over the uninteresting directions.

\begin{figure}[t!]
\centering
\vspace*{-4cm}
\includegraphics[width=0.8 \linewidth]{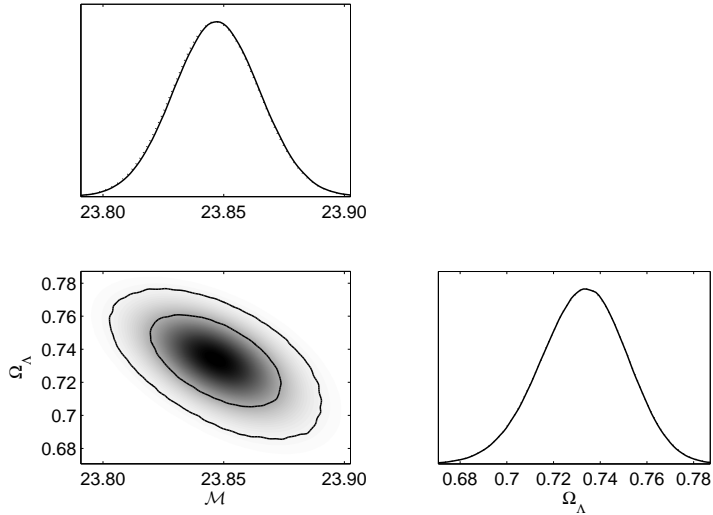}
\vspace*{-4.5cm}
\caption{Output of a very simple two-parameter MCMC analysis. The data set is a combination of supernova, CMB, and galaxy correlation data, and the two parameters are the cosmological constant $\Omega_\Lambda$ and the standardized supernova brightness ${\cal M}$. The two-dimensional distribution is in the bottom left and the marginalized single-parameter constraints in the other two boxes. In this case, both parameters have accurately gaussian distributions and are correlated. [Figure reproduced from Ref.~\cite{recquint2}.]}
\label{f:liddle2}
\end{figure}

Figure~\ref{f:liddle2} shows a typical outcome of a simple MCMC calculation.

In mapping the posterior using the point density of the chains, one in fact ignores the actual values of the probability in the chains, since the chain point locations themselves encode the posterior. However, one can also plot the posterior by computing the average likelihood in a parameter-space bin (marginalized or not), which should of course agree. Whether it does or not is an additional guide to whether the chain is properly converged.

In addition to analyzing the posterior probability from the chains, both to plot the outcome and extract confidence levels to quote as headline results, there are a number of other ways of using them. Two examples are
\begin{itemize}
\item Importance sampling. If new data becomes available, rather than computing new chains from scratch one can importance sample existing chains, by reweighting the elements according to the new likelihood \cite{LewBrid}. One can also use importance sampling to study the effect of varying the prior. These operations mean that the points now have non-integer weights, but this creates no new issue of principle. However importance sampling may result in an insufficient density of points in the new preferred region, making the sampling of the posterior noisier than would be possible with new chains.
\item Bayesian complexity. This quantity measures the number of parameters actually constrained by the data in hand \cite{Spieg,KTP}, which may be less than the number of parameters of the model, either because overall the data are poorly constraining, or because of specific parameter degeneracies leaving some parameter combinations unconstrained. It can also be used to determine a semi-Bayesian model selection statistic known as the Deviance Information Criterion, discussed briefly in Section~\ref{ss:ic}
\end{itemize}

\subsubsection{Other sampling algorithms}

Metropolis--Hastings achieves the desired goal, but may struggle to do so efficiently if it is difficult to find a good proposal function or if the assumption of  a fixed proposal function proves disadvantageous. This has tended not to be a problem in cosmological applications to date, but it is nevertheless worthwhile to know that there are alternatives which may be more robust. Some examples, all discussed in MacKay's book \cite{MacKay}, are
\begin{itemize}
\item Slice sampling: This method allows the proposal function to change during the calculation, tuning itself to an appropriate scale, though there is an additional computational cost associated with enforcing the detailed balance condition. The steps are made in a single parameter direction at a time, hence the name, and cycle through the parameter directions either sequentially or randomly. Slice sampling is implemented in CosmoMC as an alternative to Metropolis--Hastings, as are some other more specialized sampling algorithms.
\item Gibbs sampling: This relies on obtaining a proposed step by sampling from conditional probability distributions, e.g.\ to step in the $\theta_1$ direction we sample from $P(\theta_1|\theta_2)$ and vice versa. It turns out that such proposals are always accepted, enhancing the efficiency. The method can however struggle to make progress along highly correlated parameter directions, traversing the diagonal through a series of short steps parallel to the axes.
\item Hamiltonian sampling: This more sophisticated approach uses an analogy with Hamiltonian dynamics to define a momentum from derivatives of the likelihood. The momentum associated with a point enables large proposal steps to be taken along trajectories of constant `energy' and is particularly well adapted to very high dimensionality problems. See Refs.~\cite{haijin,taylor} for cosmological applications.
\end{itemize}

\subsubsection{Machine learning}

The slow likelihood evaluations, stemming mainly from the time needed to predict observational quantities such as the CMB power spectra from the models, remain a significant stumbling block in cosmological studies. One way around this may be to use machine learning to derive accurate theoretical predictions from a training set, rather than carry out rigorous calculations of the physical equations at each parameter point. Two recent attempts to do this are PICO \cite{pico} and CosmoNet \cite{cosmonet}, the former also allowing direct estimation of the WMAP likelihood from the training set. This is a promising method, though validation of the learning output when new physical processes are included may still mean that many physics-based calculations need to be done.

\subsubsection{Overview}

In conclusion, cosmological parameter estimation from data is increasingly regarded as a routine task, albeit one that requires access to significant amounts of computing resource. This is principally attributed to the public availability of the CosmoMC package, and the continued work by its authors and others to enhance its capabilities. On the theoretical side, an impressive range of physical assumptions are supported, and many researchers have the expertise to modify the code to further broaden its range. On the observational side, researchers have recognized the importance of making their data available in formats which can readily be ingested into MCMC analyses. The WMAP team were the first to take this aspect of delivery very seriously, by publically releasing the `WMAP likelihood code' at the same time as their science papers, allowing cosmologists everywhere to immediately explore the consequences of the data for their favourite models. Indeed, I would argue that by now the single most important product of an observational programme would be provision of a piece of software calculating the likelihood as a function of input quantities (e.g.\ the CMB power spectra) computable by CosmoMC.

\subsection{Uncertainties, biases and significance}

The significance with which results are endowed appears to be strongly dependent on the science community obtaining them. At one extreme lies particle physics experimentalists, who commonly set a `five-sigma' threshold to claim a detection (in principle, for gaussian uncertainties, corresponding to 99.99994\% probability that the result is not a statistical fluke). By contrast, astrophysicists have been known to get excited by results at or around the `two-sigma' level, corresponding to 95.4\% confidence for a gaussian. At the same time, there is clearly a lot of skepticism as to the accuracy of confidence limits; some possibly apocryphal quotes circulating in the data analysis community include ``Once you get to three-sigma you have about a half chance of being right.'' and ``95\% of 95\% confidence results do not turn out to be right; if anything 95\% of them turn out to be wrong''.

There are certainly good reasons to think that results are less secure than the stated confidence from a likelihood analysis would imply. Amongst these are
\begin{itemize}
\item In realistic cases, the probability may not fall as fast as a gaussian in the tails of the distribution even if accurately gaussian near the peak.
\item There may be unmodelled systematic errors. The natural trend in a maturing observational field is to be initially dominated by statistical uncertainty, but as instrumental accuracy improves to then reach a systematic floor where it becomes difficult or impossible to model extraneous physical effects (e.g.\ population evolution in supernovae which one is planning to use as standard candles over cosmic epochs).
\item The likelihood function may be uncertain in a way not included in the quoted uncertainty. For instance there are now several different treatments of the WMAP likelihood, differing in the way the beam profiles, source subtraction, or low multipole likelihoods are calculated.
\item The researchers may, consciously or otherwise, have adopted a model motivated by having seen the data, and then attempted to verify it from the same data. It has recently been claimed that this problem is widespread in the neuroscience field \cite{neuro}, with the authors of that study suspecting the issue extends to other disciplines. An example of this would be to spot an unusual feature in, say, a cosmic microwave background map, and then attempt to assess the probability of such a feature using Monte Carlo simulations. This ignores the fact that there may have been any number of other no more unlikely features, that {\it weren't seen in the data}. Consider the well-publicized appearance of Stephen Hawking's initials in the WMAP maps, obviously massively improbable {\it a priori} but nonetheless visible in Fig.~\ref{f:liddle1} (in blue, just above the middle axis, somewhat left of center). 
\item Publication bias: positive results are more likely to get published than negative ones, e.g.\ the 95\% confidence result you just read about arose from one of 20 studies, the other 19 of which generated null results. This is widely recognized in the medical statistics community,\footnote{Ioannidis \cite{Ioannidis} goes so far as to claim a proof that most published results are false, publication bias being partly responsible.} leading to introduction of costly treatments that may be ineffective or even harmful. In this context, the additional problem may exist that trials are funded by large companies whose profitability depends on the outcome, and who may be in a position to influence whether they are published.
In cosmology, this may be a particular problem for cosmic non-gaussianity studies, where many different independent tests can be carried out.
\item Model uncertainty: the possibility of different models, rather than just parameter values, describing the data has not be consistently allowed for.
\end{itemize}
I'm aware of situations where all of these have been important, and in my view 95\% confidence is not sufficient to indicate a robust result.  This appears to increasingly be the consensus in the cosmology community, perhaps because too much emphasis was put on two 95\% confidence level results in the first-year WMAP data (a high optical depth and running of the spectral index) which lost support in subsequent data releases.

On the other hand, `five-sigma' may be rather too conservative, intended to give a robust result in all circumstances. In reality, the confidence level at which a result becomes significant should depend on the nature of the model(s) under consideration, and the nature of the data obtained.  Some guidance on where to draw the line may be obtained by exploring the issue of model uncertainty, the topic of the next section.

\section{COSMOLOGICAL MODEL SELECTION}

\label{c:modsel}

\subsection{Model selection versus parameter estimation}

Estimation of cosmological parameters, as described in the previous section, assumes that we have a particular model in mind to explain the data. More commonly, however, there tends to be competing models available to describe the data, invoking parametrizations of different physical effects. Each model corresponds to a different choice of variable parameters, accompanied by a prior distribution for those parameters. Indeed, the most interesting questions in cosmology tend to be those about models, because those are the qualitative questions. Is the Universe spatially flat or not? Does the dark energy density evolve? Do gravitational waves contribute to CMB anisotropies? We therefore need techniques not just for estimating parameters within a model, but also for using data to discriminate between models. Bayesian tools are particularly appropriate for such a task, though I also describe some non-Bayesian alternatives at the end of this section. A comprehensive review of Bayesian model selection as applied to cosmology was recently given by Trotta \cite{Trotrev}.

An important implication of model-level Bayesian analysis is that there is a clear distinction between a model where a quantity is fixed to a definite value, versus a more general model where that parameter is allowed to vary but happens to take on that special value. A cosmological example is the dark energy equation of state, $w$, which the cosmological constant model predicts (in good agreement with current observations) to be precisely $-1$, and which other models such as quintessence leave as a free parameter to be fit from data. Even if it is the cosmological constant which is the true underlying model, a model in which the equation of state can vary will be able to fit any conceivable data just as well as the cosmological constant model (assuming of course that its range of variation includes $w=-1$). What distinguishes the models is {\em predictiveness}.

As an example, consider a magician inviting you to ``pick a card, any card''. Since you don't know any better, your `model' tells you that are equally likely to pick any card. The one you actually do pick is not particularly surprising, in the sense that whatever it was, it was compatible with your model. The magician, however, has a much more predictive model; by whatever means, they know that you will end up picking the queen of clubs. Both models are equally capable of explaining the observed card, but the magician's much more predictive model lets them earn a living from it. Note in this example, any surprise that you might feel comes not from the card itself, but from the magician's ability to predict it. Likewise, a scientist might find themselves surprised, or dismayed, as incoming data continue to lie precisely where some rival's model said they would.

Model selection/comparison is achieved by choosing a ranking statistic which can be computed for each model, allowing them to be placed in rank order. Within the Bayesian context, where everything is governed by probabilities, the natural choice is the model probability, which has the advantage of having a straightforward interpretation.

\subsection{The Bayesian evidence}

The extension of the Bayesian methodology to the level of models is both unique and straightforward, and exploits the normalizing factor $P(D)$ in equation (\ref{e:bayes2}) which is irrelevant to
and commonly ignored in parameter estimation. 

We now assume that there are several models on the table, each with their own probability $P(M_i)$ and explicitly acknowledge that our probabilities
are conditional not just on the data but on our assumed model $M$,
writing
\begin{equation}
\label{e:evparm}
P(\theta|D,M) = \frac{P(D|\theta,M)P(\theta|M)}{P(D|M)} \,.
\end{equation}
This is just the previous equation with a condition on M written in each term.
The denominator, the probability of the data given the model, is by
definition the model likelihood, also known as the \emph{Bayesian
evidence}.  Note that, unlike the other terms in this equation, it does not depend on specific values for the parameters $\theta$ of the model.

The evidence is key because it appears in yet another rewriting of Bayes theorem, this time as
\begin{equation}
P(M|D) = \frac{P(D|M)P(M)}{P(D)} \,.
\end{equation}
The left-hand side is the posterior model probability (i.e.\ the probability of the model given the data), which is just what we want for model selection. To determine it, we need to compute the Bayesian
evidence $P(D|M)$, and we need to specify the prior model probability $P(M)$. It is a common convention to take the prior model probabilities to be equal (the model equivalent of a flat parameter prior), but this is by no means essential.

To obtain an expression for the evidence, consider
Eq.~(\ref{e:evparm}) integrated over all $\theta$. Presuming we have
been careful to keep our probabilities normalized, the left-hand side
integrates to unity, while the evidence on the denominator is
independent of $\theta$ and comes out of the integral. Hence
\begin{equation}
\label{e:evidence}
P(D|M) = \int P(D|\theta,M)P(\theta|M) d\theta \,,
\end{equation}
or, more colloquially,
\begin{equation}
\mbox{Evidence} = \int \left( \mbox{Likelihood} \, \times \,
\mbox{Prior} \right) \,  d\theta \,.
\end{equation}
In words, the evidence $E$ is the average likelihood of the parameters
averaged over the parameter prior. For the distribution of parameter
values you thought reasonable before the data came along, it is the
average value of the likelihood.

The Bayesian evidence rewards model predictiveness. For a model to be
predictive, observational quantities derived from it should not depend
very strongly on the model parameters. That being the case, if it fits
the actual data well for a particular choice of parameters, it can be
expected to fit fairly well across a significant fraction of its prior
parameter range, leading to a high average likelihood. An unpredictive
model, by contrast, might fit the actual data well in some part of its
parameter space, but because other regions of parameter space make
very different predictions it will fit poorly there, pulling the
average down. Finally, a model, predictive or otherwise, that cannot
fit the data well anywhere in this parameter space will necessarily
get a poor evidence.

Often predictiveness is closely related to model simplicity; typically
the fewer parameters a model has, the less variety of predictions it
can make. Consequently, model selection is often portrayed as
tensioning goodness of fit against the number of model parameters, the
latter being thought of as an implementation of Ockham's
razor. However the connection between predictiveness and simplicity is
not always a tight one. Consider for example a situation where the
predictions turn out to have negligible dependence on one of the
parameters (or a degenerate combination of parameters). This is
telling us that our observations lack the sensitivity to tell us
anything about that parameter (or parameter combination). The
likelihood will be flat in that parameter direction and it will
factorize out of the evidence integral, leaving it unchanged. Hence
the evidence will not penalize the extra parameter in this case,
because it does not change the model predictiveness.

The ratio of the evidences of two models $M_0$ and $M_1$ is known as
the Bayes factor \cite{KR95}:
\begin{equation}
B_{01} \equiv \frac{E_0}{E_1} \,,
\end{equation}
which updates the prior model probability ratio to the posterior
one. Some calculational methods determine the Bayes factor of two
models directly. Usual convention is to specify the logarithms of the
evidence and Bayes factor.

\subsection{Calculational methods}

Equation (\ref{e:evidence}) tells us that to get the evidence for a model, we need to integrate the likelihood throughout the parameter space. In principle this is a very standard mathematical problem, but it is made difficult because the integrand is likely to be extremely highly peaked and we do not know in advance where in parameter space the peak might be. Further, the parameter space is multi-dimensional (between about 6 and 10 dimensions would be common in cosmological applications), and as remarked in Section~\ref{c:cospar} the individual likelihood evaluations of the integrand at a point in parameter space are computationally expensive (a few CPU seconds each), limiting practical calculations to $10^5$ to $10^6$ evaluations. Successful Bayesian model selection algorithms are therefore dependent on efficient algorithms for tackling this type of integral.

Model probabilities are meaningful in themselves and don't require further interpretation, but it is useful to have a scale by which to judge differences in evidence. The usual scale employed is the Jeffreys' scale \cite{Jeff} which, given a difference $\Delta \ln E$ between the
evidences $E$ of two models, reads
\begin{center}
\begin{tabular}{|c|l|}
\hline
$\Delta \ln E < 1$ & Not worth more than a bare mention.\\
$1 < \Delta \ln E < 2.5$ & Significant.\\
$2.5 < \Delta \ln E < 5$ & Strong to very strong.\\
$5 < \Delta \ln E$ & Decisive.\\
\hline
\end{tabular}
\end{center}
In practice the divisions at 2.5 (corresponding to posterior odds of about 13:1) and 5 (corresponding to posterior odds of about 150:1) are the most useful. 

As the main steps in the Jeffreys' scale are 2.5, this sets a target accuracy in an evidence calculation; it should be accurate enough that calculational uncertainties do not move us amongst these different categories. The accuracy should therefore be better than about $\pm 1$. However there is no point in aiming for an accuracy very much better than that, which would not tell us anything further about the relative merits of the models. Hence an accuracy of a few tenths in $\ln E$ is usually a good target.

\subsubsection{Exact methods}

\begin{figure}[th!]
\begin{center}
\includegraphics[width=0.7 \linewidth]{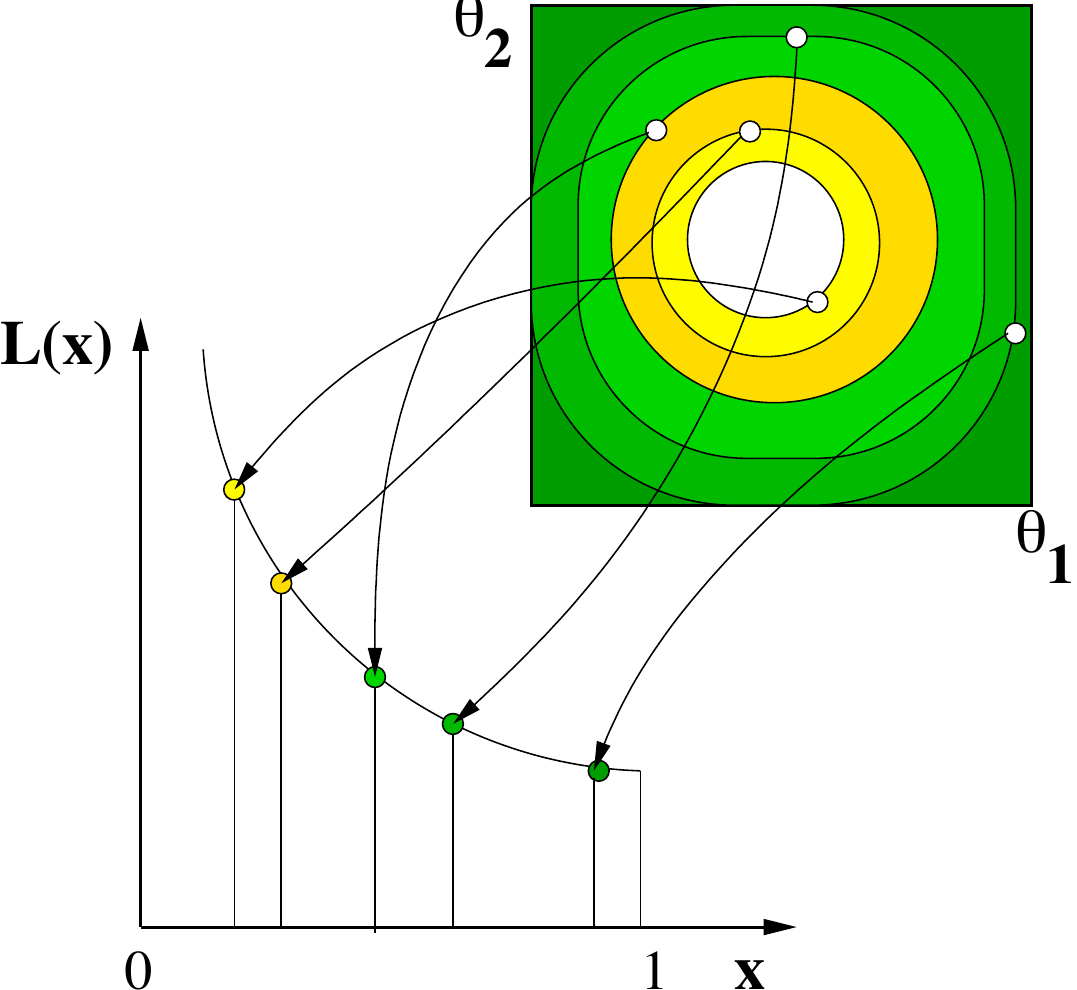}
\caption{A schematic of the nested sampling algorithm. The
two-dimensional parameter space is shown at the top right. The points
within it are considered to represent contours of constant likelihood,
which sit within each other like layers of an onion (there is however
no need for them to be simply connected). The volume corresponding to
each thin shell of likelihood is computed by the algorithm, allowing
the integral for the evidence to be accumulated as shown in the
graph. [Figure courtesy David Parkinson.]}
\label{fig:nested}
\end{center}
\end{figure}

Computing the evidence is more demanding than mapping the dominant part of the posterior, as we now have to be able to accurately integrate the likelihood over the entire parameter space. There are several methods which become exact in the limit of infinite computer time, and are capable of offering the desired accuracy with finite resources. Those used so far in cosmological settings are
\begin{itemize}
\item Thermodynamic integration. This variant on Metropolis--Hastings MCMC varies the effective temperature of the chain, allowing it to explore lower likelihood regions and hence fully probe the prior space. While well regarded in the statistics community, cosmological applications to date have proven extremely CPU demanding, e.g.\ Ref.~\cite{Beltran}.
\item Nested sampling. Introduced by Skilling \cite{Skilling}, this algorithm explores parameter space with a large collection of points, typically hundreds. The lowest likelihood point is deleted and replaced by a randomly-drawn point of higher likelihood, the cluster of points in this way migrating to the high likelihood regions as shown in Figure~\ref{fig:nested}. As a byproduct it generates a set of posterior samples that are suitable for parameter estimation. It was first implemented for cosmology in Refs.~\cite{MPL,PML}, though effectively limited to cases with a  single strong likelihood peak. A more powerful implementation, MultiNest \cite{FHB} is capable of handling multi-peaked likelihoods. Nested sampling is currently the method of choice in cosmology and has also been applied to supersymmetry \cite{feroz}.
\item VEGAS. This is a multi-dimensional integrator much used by particle physicists \cite{lepage}. It has only been used once  in cosmological model selection thus far \cite{SHM} but shows promise as an alternative to nested sampling.
\end{itemize}

\subsubsection{Approximate or restricted methods}

In addition to brute force numerical methods to compute the evidence, various approximate or restricted methods exist which may be useful in some circumstances. The most important three, discussed in more detail in Refs.~\cite{Trotrev,lid04}, are
\begin{enumerate}
\item The Laplace approximation: one assumes that the likelihood is adequately described by a multi-variate gaussian, expands around the maximum and carries out the evidence integral analytically. This may work well at least in the absence of strong degeneracies, but it can be hard to assess its accuracy.
\item The Savage--Dickey ratio: This computes the Bayes factor for the case where one model is nested within another. It amounts to a careful evaluation of the marginalized posterior of the more complex model, evaluated at the parameter value of the embedded model. It is in principle exact and can be estimated from Markov chains. However, if the embedded model is not near the actual maximum likelihood, so that one is trying to judge whether the simple model is excluded (likely to be the case most of interest), there may be few Markov chain samples near the embedded model and the estimate becomes noisy. For this reason a general implementation of the Savage--Dickey method has not yet been made available, though it has been successfully applied in specific examples, e.g.\ Ref.~\cite{Trotta07a}.
\item The Bayesian Information Criterion (BIC): Introduced by Schwarz \cite{Schwarz}, this approximates the Bayes factor without requiring the models to be nested, using a Laplace approximation and further assumptions on the nature of the data. The BIC is given by
\begin{equation}
\label{e:BIC}
{\rm BIC} = -2 \ln {\cal L}_{\rm max} +k \ln N \,,
\end{equation}
where ${\cal L}_{\rm max}$ is the maximum likelihood achievable by the model,  $k$ the number of parameters, and $N$ the number of datapoints. Models are ranked according to their BIC values. Despite the name, it has nothing to do with information theory, but rather was named by analogy to information theory measures discussed in Section~\ref{ss:ic}. It has been used quite widely in cosmology, but the validity of the approximation needs careful consideration and full evidence evaluation is to be preferred whenever possible.
\end{enumerate}

\subsection{Multi-model inference}

Ultimately, a successful implementation of model selection should result in all models being eliminated except for a single survivor. This model will encode all the physics relevant to our observations, and we can then proceed to estimate the parameters of the surviving model to yield definitive values of quantities of interest, e.g.\ the baryon density, neutrino masses, etc. Unfortunately, in practice it is quite likely that more than one model might survive in the face of the data, as has always been the case so far in any analysis I've done. Despite that, we might well want to know what the best current knowledge of some of the parameters is.

Fortunately, the Bayesian framework again supplies a unique procedure for extracting parameter constraints in the presence of model uncertainty. One simply combines the probability distribution of the parameters within each individual model, weighted by the model probability. This is very much analogous to quantum mechanics, the superposition of model states being equivalent to a superposition of eigenstates of an observable such as position. The combination is particularly straightforward if one already has a set of samples from the likelihood, such as a Markov chain, for each model; we then just concatenate the chains giving each point a fractional weighting according to the model probability of the chain it came from.

Note that in this superposition, some parameters may have fixed values in some of the models. For example, if we are interested in the dark energy equation of state $w$, we will no doubt be including a cosmological constant model in which $w$ is fixed to $-1$. Within this model, the probability distribution is an appropriately normalized delta function, and the combined probability distribution is therefore neither continuous nor differentiable. This can have the interesting consequence that some confidence limits may have zero parameter width. See Ref.~\cite{LMPW} for some actual calculations.

\subsection{Other approaches to model selection}

\label{ss:ic}

Within the Bayesian framework, the evidence is the natural model selection statistic. However, other paradigms suggest alternatives. The most prevalent is information theory approaches, where the statistics are known as {\it information criteria}. The first and most widespread is the Akaike Information Criterion (AIC), defined by \cite{Akaike}
\begin{equation}
\label{e:AIC}
{\rm AIC} = -2 \ln {\cal L}_{\rm max} +2k \,,
\end{equation}
using the same terminology as equation (\ref{e:BIC}).
This sets up a tension between goodness of fit and complexity of the model, with the best model minimizing the AIC. There is also a slightly modified version applicable to small sample sizes, which ought to be used in any case \cite{BA02}. It is not always clear how big a difference in AIC is required for the worse model to be significantly disfavoured, but a typical guide (actually obtained by attempting a Bayesian-style probabilistic interpretation) is a difference of 6 or more should be taken seriously.

Beyond the AIC, there are a large variety of different information criteria in use.
Burham and Anderson \cite{BA02} give an excellent textbook account of information theory approaches to model selection, and cosmological applications of some of them are discussed in Refs.~\cite{lid04,lid07}. 

Perhaps most worthy of special mention is the Deviance Information Criterion of Spiegelhalter et al.~\cite{Spieg}, which combines aspects of the Bayesian and information theory approaches. It
replaces the $k$ in equation (\ref{e:AIC}) with the Bayesian complexity mentioned towards the end of Section~\ref{ss:mcmeth} (for technical reasons it also replaces the maximum likelihood with the likelihood at the mean parameter values, but this is not so significant). The complexity measures the number of model parameters actually constrained by the data. In doing so it overcomes the main difference between the Bayesian and Information Criterion approaches, which is their handling of parameter degeneracies. The Information Criterion approach penalizes such parameters, but the Bayesian method does not as the integral of the likelihood over the degenerate parameter factors out of the evidence integral. The Bayesian view is that in such cases the more complex model should not be penalized, because the data are simply not good enough to say whether the degenerate parameter is needed or not. As this seems more reasonable, the DIC may be preferable to the AIC.

An alternative model selection approach comes from algorithmic information theory, and has variants known as minimum message length and minimum description length \cite{wallace}. These interpret the best model as being the one which offers the maximal compression of the data, i.e.\ that the combination of model plus data can be described in the smallest number of bits. As a concept, this idea remarkably originates with Leibniz in the 17th century.

\section{FORECASTING AND EXPERIMENTAL DESIGN}

The above discussion concerned analysis of data which was in hand. Another important application of statistical techniques is to forecast the outcomes of future experiments, which is increasingly expected by funding agencies wishing to compare the capabilities of competing proposals. A Figure of Merit (FoM) is defined for each experiment, and used to rank experiments. More ambitiously, the same techniques can be used to optimize the design of a survey in order to maximize the likely science return (e.g.\ Ref.~\cite{surveys}).

\subsection{Fisher matrix approaches}

The leading approach presently is the Fisher information matrix approach, introduced to cosmology in Ref.~\cite{fisher} and popularized especially when adopted by the initial report of the Dark Energy Task Force \cite{DETF}. The Fisher matrix measures the second derivatives of the likelihood around its maximum, and in a gaussian approximation is used to estimate the expected uncertainty in parameters around some selected fiducial model. For instance, the DETF FoM considers the uncertainty on the two parameters of the dark energy equation of state defined by the form $w = w_0 +(1-a)w_a$, taking the fiducial model to be $\Lambda$CDM ($w_0=-1$, $w_a = 0$), defining the FoM to be the inverse area of the 95\% confidence region. The likelihood is determined via a model of how well the instrument will perform, possibly through analysis of a simulated data stream.

An important caveat is the following: an experiment capable of reducing the volume of permitted parameter space by say a factor of 10 should in no way be considered as having a 90\% chance of measuring those parameters as different from some special value. An example is whether upcoming data can exclude the cosmological constant model in favour of dark energy. What the Fisher FoM shows is the reduction in allowed parameter volume {\it provided that the dark energy model is correct}. However that assumption is actually what we wish to test; it is perfectly possible that it is the cosmological constant model which is correct and then no amount of improvement to the uncertainty will rule it out. Once again, to develop a full picture we need to consider multiple models, and hence model selection statistics.

\subsection{Model selection approaches}

Model selection forecasting, pioneered for cosmology in Refs.~\cite{Trotta07a,MPCLK}, instead forecasts the ability of upcoming experiments to carry out model comparisons. Let's restrict to the simplest case of two models, one nested within another, though the generalization to other circumstances is straightforward.

The very simplest model selection question is to assume that the nested model (say $\Lambda$CDM) is true, and ask whether a given experiment would be able to rule out the more complex alternative. Data is simulated only for the $\Lambda$CDM model, as with the Fisher analysis, and the Bayes factor between $\Lambda$CDM and the parametrized dark energy model calculated. This shows how strongly the experiment will rule out the dark energy model if it is wrong.

More generally, one may wish to consider the dark energy model as the correct one, and ask whether the simpler model can be excluded. This is more complex, as the outcome depends on the actual parameters of the dark energy model (known as the fiducial parameters), and hence has to be considered as a function of them. This leads to the concept of the Bayes factor plot, showing the expected Bayes factor as a function of fiducial parameters \cite{MPCLK}. A suitable FoM may be to minimize the area in fiducial parameter space in which the wrong model cannot be ruled out by the proposed experiment. Alternatively, one can study the distribution of the Bayes factor weighted by present knowledge of the parameters, to predict the probability distribution of expected outcomes of the experiment \cite{Trotta07a}.

Examples of cosmological model selection forecasts can be found in Refs.~\cite{Trotta07a,MPCLK,LMPW}.

\section{THE END}

I was going to write a brief summary. Decided not to.

\section*{Acknowledgments}

This work was supported by STFC (UK). I thank my collaborators on these topics --- Maria Beltran, Pier Stefano Corasaniti, Marina Cort\^es, Juan Garcia-Bellido, Martin Kunz, Ofer Lahav, Samuel Leach, Julien Lesgourgues, Pia Mukherjee, C\'edric Pahud, David Parkinson, Martin Sahl\'en, Anze Slosar, Roberto Trotta, and Yun Wang --- for numerous discussions, with special appreciation for Pia and David's work with me on the nested sampling method. I also thank Sarah Bridle, George Efstathiou, Alan Heavens, Mike Hobson, Andrew Jaffe, Eric Linder, and John Peacock for many debates concerning Bayesian data analysis.


\section*{LIST OF ACRONYMS}

\begin{tabular}{ll}
AIC & Akaike information criterion \\
BIC & Bayesian information criterion \\
CDM & Cold dark matter \\
CMB & Cosmic microwave background\\
CosmoMC & [A software package for MCMC analysis]\\
DETF & Dark energy task force\\
FoM & Figure of merit\\
$\Lambda$CDM & Lambda cold dark matter (model)\\
MCMC & Markov Chain Monte Carlo\\
WMAP & Wilkinson Microwave Anisotropy Probe
\end{tabular}

\end{document}